\documentclass[aps,prl,twocolumn,amsmath,amssymb,nofootinbib,
superscriptaddress,floatfix]{revtex4}
\usepackage{amssymb}
\usepackage{color}
\usepackage{graphicx}
\usepackage{bm}
\usepackage{subfigure}

\def\avg#1{\left\langle#1\right\rangle}
\def\bra#1{\left\langle#1\right|}
\def\ket#1{\left|#1\right\rangle}
\def\braket#1#2{\left\langle #1\right|\!\left.#2\right\rangle}

\def\sgn{{\rm sgn}}

\def\be{\begin{equation}}       \def\ee{\end{equation}}
\def\bea{\begin{eqnarray}}      \def\eea{\end{eqnarray}}
\def\ba{\begin{array} }
\def\ea{\end{array} }
\def\bnum{\begin{enumerate} }
\def\enum{\end{enumerate}}

\def\nn{\nonumber}

\def\=>{\Rightarrow}
\def\>{\rightarrow}
\def\A{\uparrow}
\def\V{\downarrow}

\def\eye2{Fathbb{I}}

\def\Eq#1{Eq.~(\ref{#1})}

\begin{document}

\title{Frustrated Resonating Valence Bond States in Two Dimensions:
Classification and Short-Range Correlations}
\author{Fan Yang}
\affiliation{School of Physics, Beijing Institute of Technology,
Beijing, 100081, China}
\author{Hong Yao}
\affiliation{Institute for Advanced Study, Tsinghua University,
Beijing, 100084, China} \affiliation{Department of Physics,
Stanford University, Stanford, California 94305, USA}

\begin{abstract}
Resonating valence bond (RVB) states are of crucial importance in
our intuitive understanding of quantum spin liquids in 2D. We
systematically classify short-range bosonic RVB states into
symmetric or nematic spin liquids by examining their flux
patterns. We further map short-range bosonic RVB states into
projected BCS wave functions, on which we perform large-scale
Monte Carlo simulations without the minus sign problem. Our
results clearly show that both spin and dimer correlations decay
exponentially in all the short-range {\it frustrated}
(non-bipartite or $Z_2$) bosonic RVB states we studied, indicating
that they are gapped $Z_2$ quantum spin liquids. Generically, we
conjecture that {\it all} short-range frustrated bosonic RVB
states in 2D have only short-range correlations.
\end{abstract}
\date{\today}
\maketitle

{\bf Introduction}: Quantum spin liquids are exotic insulators
which cannot be adiabatically connected into a band insulator and
which can support fractionalized excitations\cite{balents}.
Introduced by Anderson nearly four decades ago\cite{anderson-73},
the resonating valence bond (RVB) state on the triangular lattice
is the first example of quantum spin liquids in more than one
dimension. Since then, there has been keen interest in searching
for such exotic states of matter in real materials as well as in
microscopic models, especially after exciting connections between
quantum spin liquids and the mechanism of high temperature
superconductivity were
suggested\cite{anderson-87,kivelson-87,lnw-06}.

%{\color{red}Although the ground state of the Heisenberg Hamiltonian for the triangular or square lattice is a Neel state instead of an RVB state, the recent numerical simulations for the Kagome lattice\cite{jiang-08,yan-11,depenbrock-12} and $J_{1}-J_{2}$ square lattice\cite{jiang-11,wang-11} report convincing evidence of the existence of fully gapped spin liquids, both of which are believed to be in the same class of short-range RVB states}.
Recently, there has been a surge of numerical simulations on
simple models reporting convincing evidence of the existence of
fully gapped spin
liquids\cite{meng-10,jiang-08,yan-11,jiang-11,wang-11,depenbrock-12},
all of which are believed to be in the same class of short-range
bosonic RVB states. Nonetheless, the nature of short-range bosonic
RVB states on various frustrated lattices has not been explicitly
revealed\cite{moessner-01,fradkin-01}, mainly because of the
so-called minus sign problem in Monte Carlo (MC) simulations of
those bosonic short-range RVB states with frustration. Because of
their conceptual importance in pictorially understanding quantum
spin liquids and their direct relevance in recent numerical
simulations, it is highly desired to unambiguously demonstrate the
nature of these short-range bosonic RVB states.

In this Letter, we systematically classify short-range bosonic RVB
states by examining their flux patterns \cite{note1}. For
instance, for the Kagome lattice we establish that there are only
four symmetric RVB states when considering only nearest-neighbor
(NN) valence bonds, as is shown in Fig. 1. Then, we show that these
bosonic short-range RVB states (including a class of RVB states with valence bonds longer than NN) can be exactly mapped into
projected BCS wave functions\cite{read-88,sorella-06,sorella-09}
on which we perform large-scale MC simulations without the minus
sign. For frustrated short-range RVB states, our simulations on
corresponding projected BCS states convincingly show that both
their spin and dimer correlations decay exponentially, indicating
that they are fully gapped $Z_2$ spin
liquids\cite{wen-91,read-91}.

\begin{figure}[tb]
\subfigure{\includegraphics[scale=0.19]{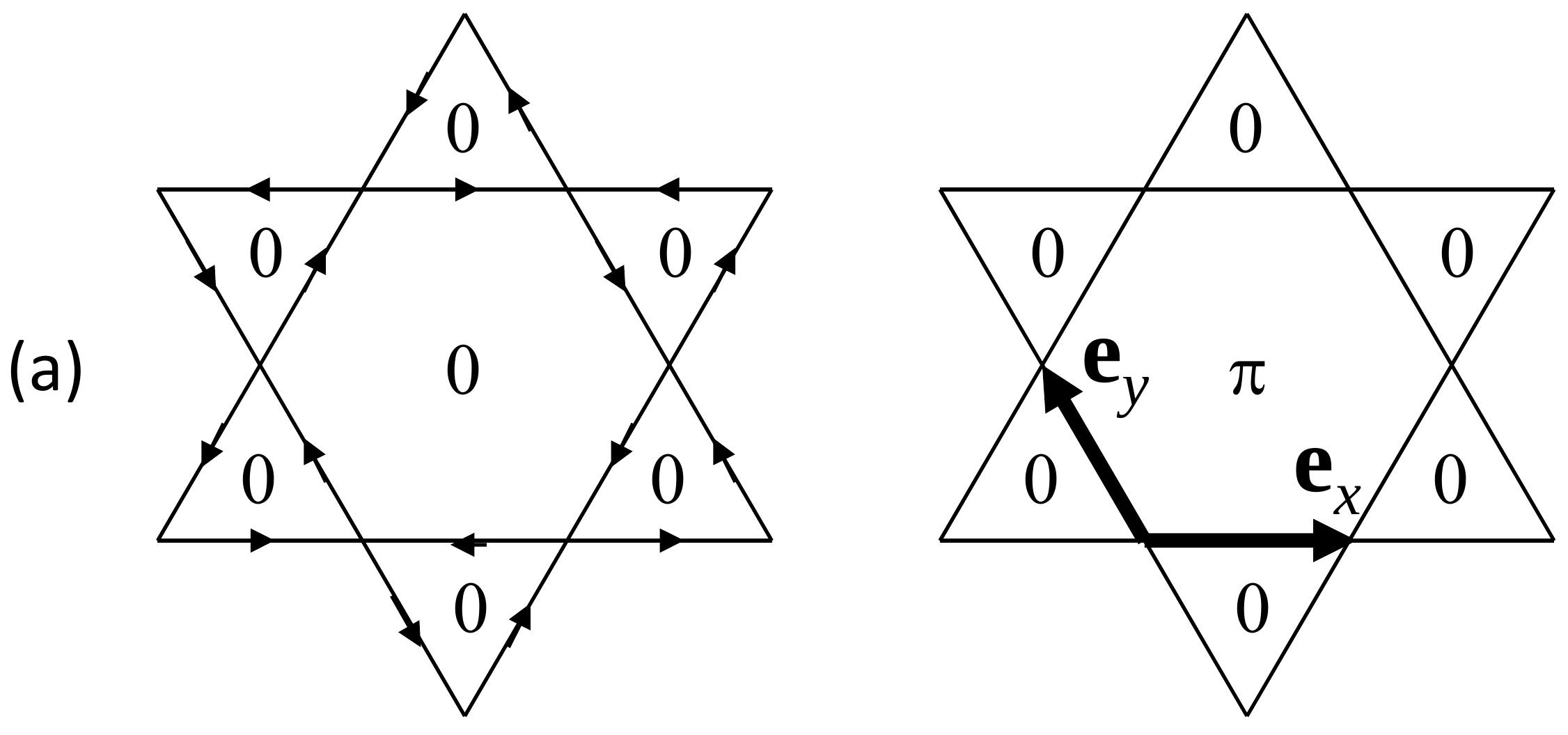}}~~
\subfigure{\includegraphics[scale=0.19]{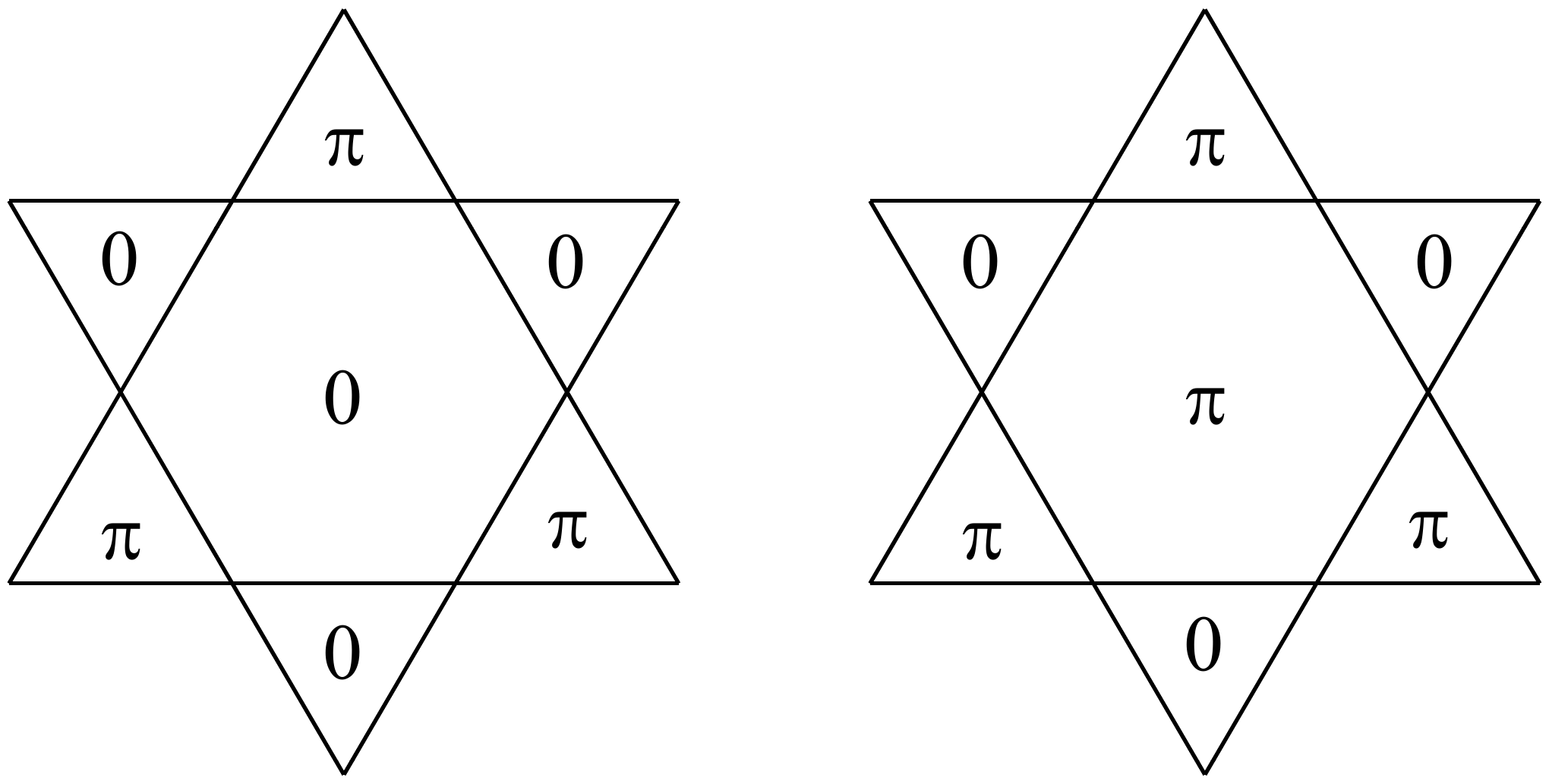}}
\subfigure{\includegraphics[scale=0.19]{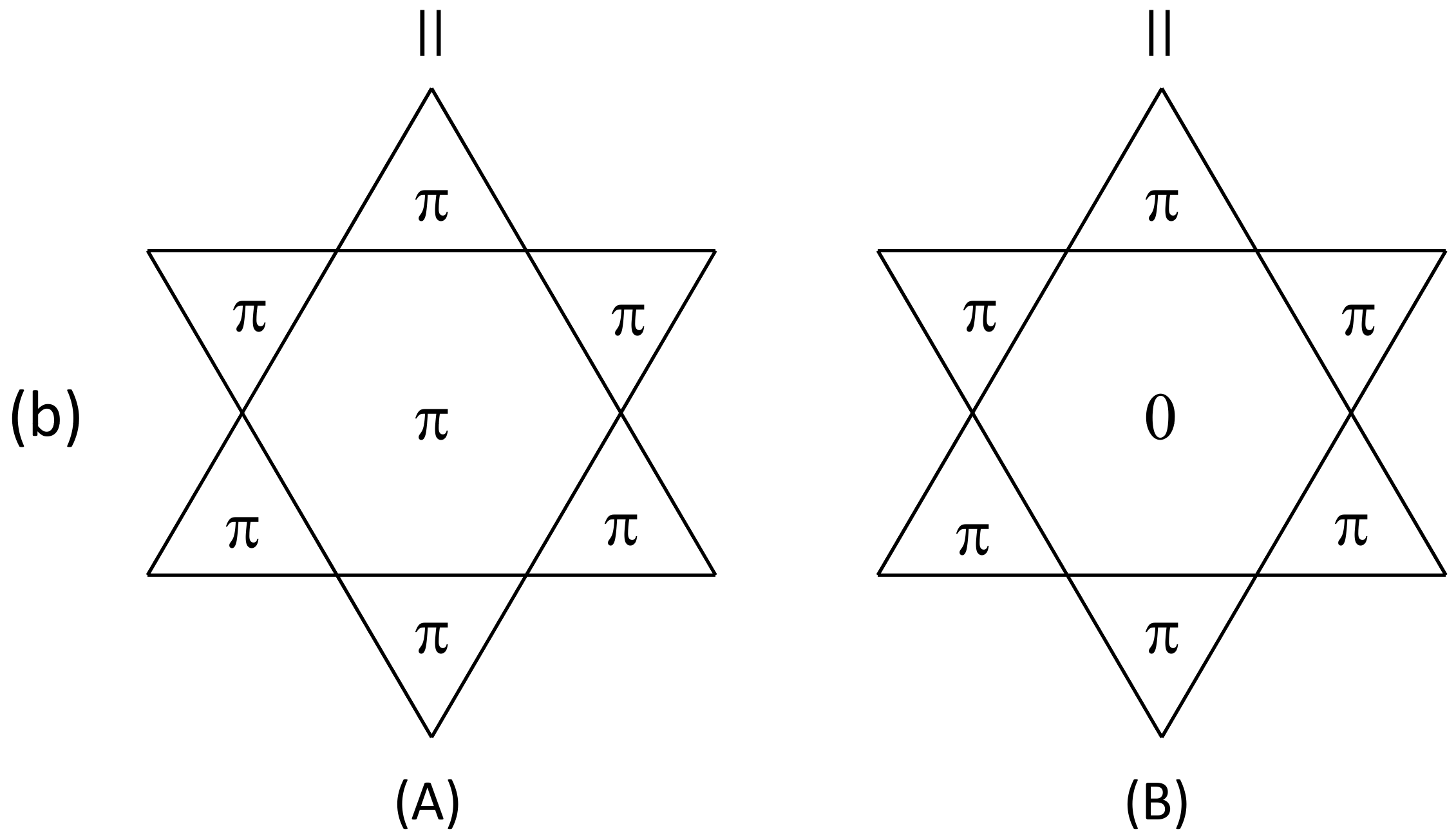}}~~
\subfigure{\includegraphics[scale=0.19]{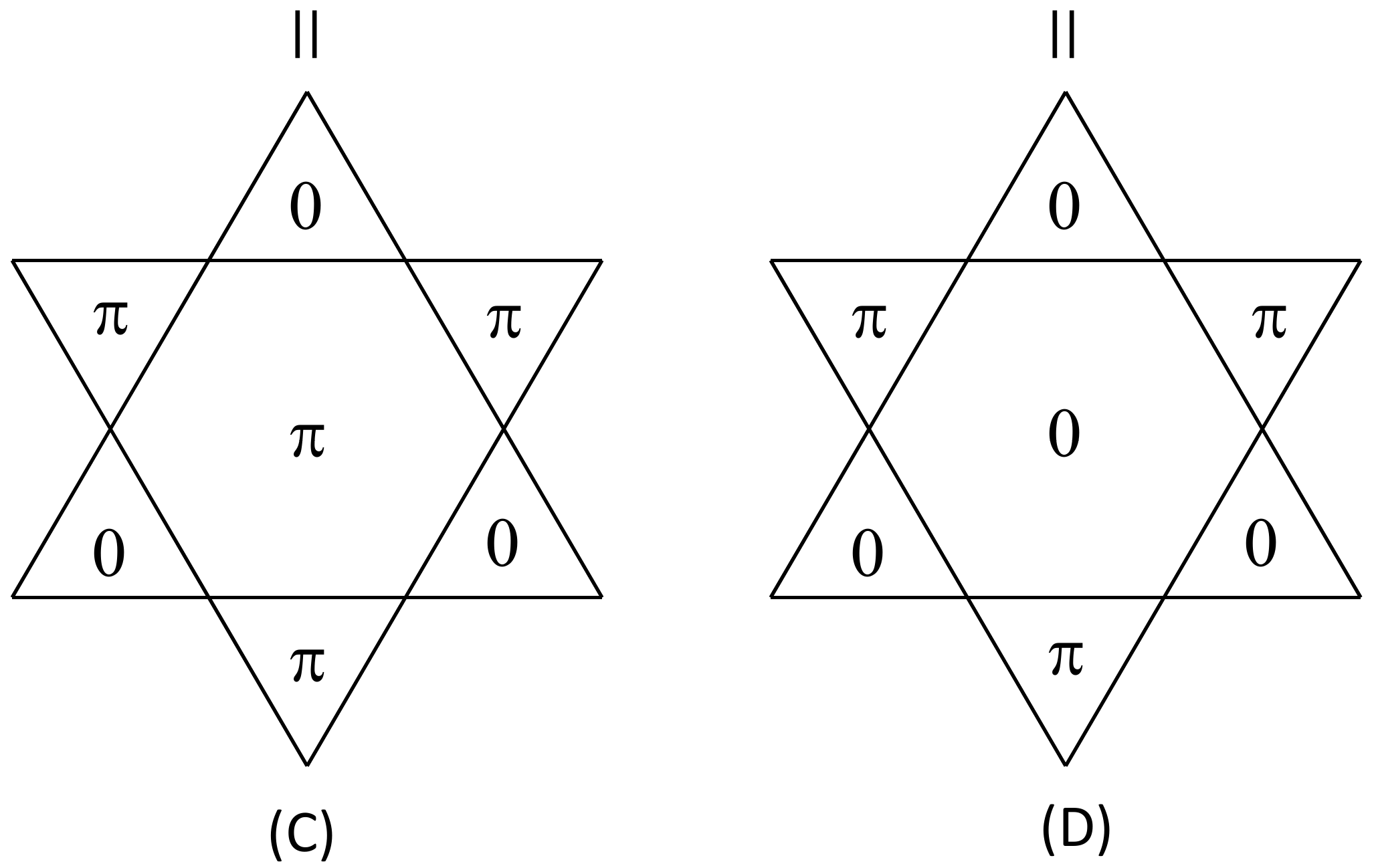}}
\caption{(a) The flux patterns $\{\phi_p\}$ of the only
 four symmetric NN-RVB states on the Kagome lattice.
 Here ${\bf e}_{x}$ and ${\bf e}_{y}$ represent the unit
 vectors.
 (b) The flux patterns $\{\phi^f_p\}$ in the corresponding projected BCS states. } \label{fig:kag}
\end{figure}

{\bf Bosonic RVB states}:  We consider the following bosonic RVB
states with NN and possibly next nearest-neighbor (NNN) valence
bonds
\bea\label{eq:brvb} \ket{\psi_\textrm{RVB}}=\sum_c
\ket{c},~~ \ket{c}= (-1)^{\delta_{c}}\prod_{(ij)\in c}
f_{ij}\ket{ij},
\eea
where $c$ labels valence bond configurations and
$\delta_c$ represents the number of bond-crossings in $c$ [the
factor $(-1)^{\delta_{c}}$ is nontrivial only for RVB states with
valence bonds beyond nearest-neighbor sites.] Here $\ket{ij}\equiv
(\ket{\A_{i}\V_{j}}-\ket{\V_{i}\A_{j}})/\sqrt2$ is the spin-singlet wave
function (or valence bond) on $(ij)$ and we assume $|f_{ij}|$ to respect all the
lattice symmetries. Note that \Eq{eq:brvb} represents a
``bosonic'' RVB state in the sense that
$\ket{\A_{i}\V_{j}}=\ket{\V_{j}\A_{i}}$. Since
$\ket{ij}=-\ket{ji}$, it is sufficient to consider $f_{ij}=-f_{ji}$. The wave
function in \Eq{eq:brvb} possesses a gauge symmetry:
$\ket{\psi_{\textrm{RVB}}}$ is invariant, up to a phase, under the
transformation $f_{ij}\to e^{i \alpha_i} f_{ij} e^{i\alpha_j}$. %On even-length plaquettes, the following defined fluxes are gauge invariant $\exp(i\phi_p)=\prod_{n\in even} \frac{f_{i_n i_{n+1}}}{|f_{i_n i_{n+1}}|}\frac{f^\ast_{i_{n+1}i_{n+2}}}{|f_{i_{n+1}i_{n+2}}|}$.
In the following, we focus on time reversal {\it invariant} RVB
states for which all $f_{ij}$ can be chosen as real; signs of
$f_{ij}$ can be represented by oriented arrows on graphs: an arrow
pointing from $i$ to $j$ means that $f_{ij}>0$, as shown in Fig.
\ref{fig:kag}. We further define flux $\phi_p=0,\pi$ (mod $2\pi$)
for plaquette $p$ through \bea \prod^{cc}_{(jk)\in
p}\sgn(f_{jk})=\exp(i\phi_p), \eea where $cc$ means that the
counterclockwise order of $(jk)$ is taken in the product above and
$\sgn$ is the sign function. It is clear that $\phi_p$ is gauge
invariant for even-length plaquette $p$. However, the gauge
transformation with $\exp(i\alpha_j)=i$ on every site $j$ changes
$\phi_p$ to $\phi_p+\pi$ for all odd-length plaquette $p$. In
other words, the two wave functions with flux patterns
$\{\phi_p\}$ and $\{\phi_p+(-1)^{n_p} \pi\}$ ($n_p$ is the length
of plaquette $p$) actually represent the same state \cite{note2}.

\begin{figure}[tb]
\includegraphics[scale=0.3]{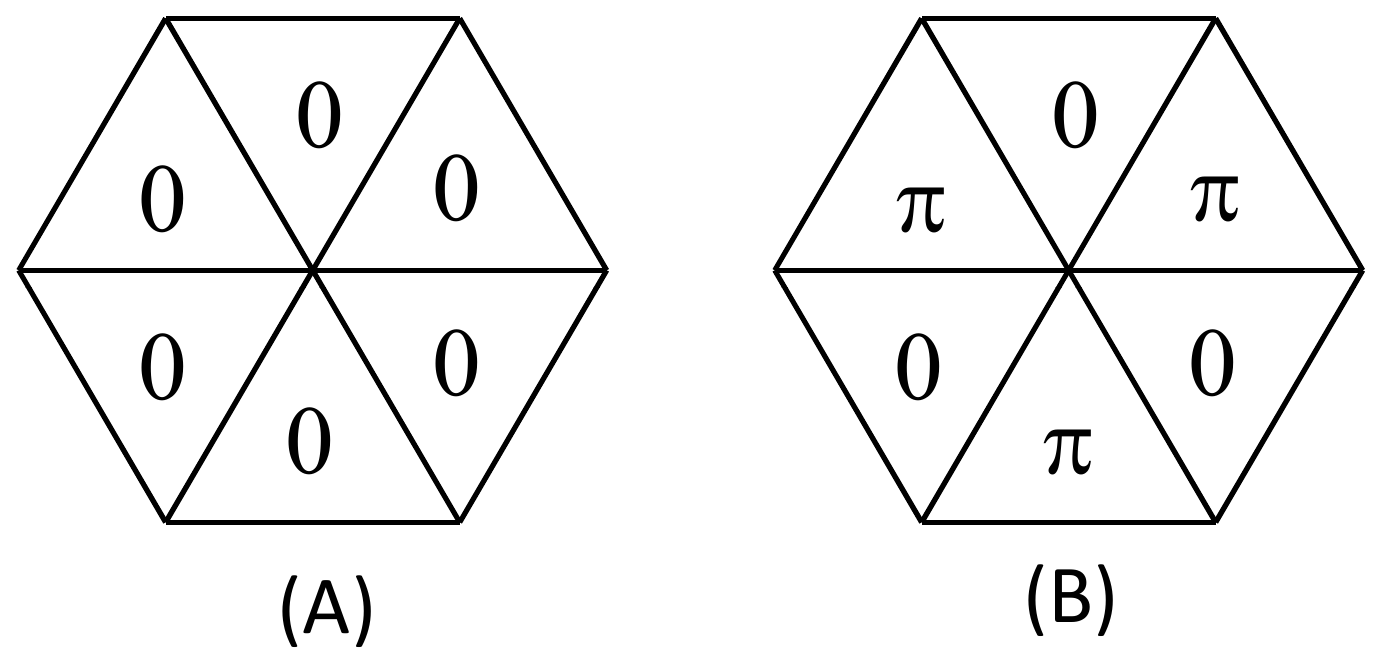}
\caption{The flux patterns $\{\phi_p\}$ of the two symmetric NN-RVB states on the triangular lattice.}
\label{fig:tri}
\end{figure}

\begin{figure}[t]
\includegraphics[scale=0.30]{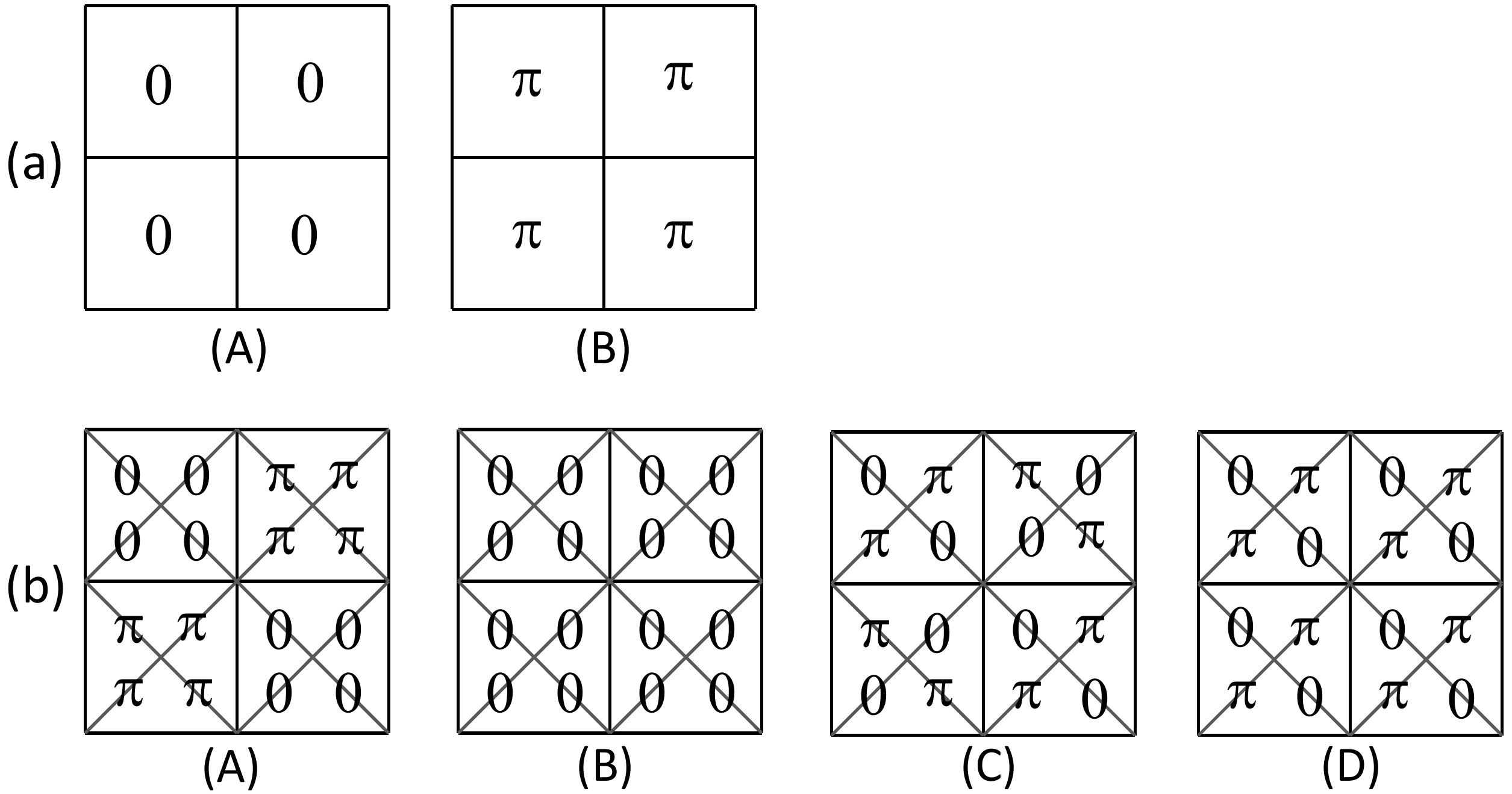}
\caption{The flux patterns $\{\phi_p\}$ of (a) the two symmetric
NN-RVB states and (b) the four symmetric NNN-RVB states (b) on the
square lattice. } \label{fig:squ}
\end{figure}

There are two questions to be answered concerning the wave
function in \Eq{eq:brvb}. First, is it a symmetric spin liquid
respecting all the symmetries of the lattice in question? Second,
do various correlations decay in power law or exponentially? The
first question can be answered by examining its flux pattern
$\{\phi_p\}$. If the flux pattern $\{\phi_p\}$ is invariant up to
the addition of $\{(-1)^{n_p}\pi\}$, under all lattice symmetry
transformations such as translations, rotations, and reflections,
the corresponding RVB state is then a symmetric spin-liquid state
. We label RVB states whose longest valence bonds are between NN
(NNN) sites as NN-RVB (NNN-RVB) states. On the Kagome lattice, we
identify four NN-RVB states as symmetric spin liquids, as shown in
Fig. \ref{fig:kag}. On the triangular lattice, only two symmetric
NN-RVB states are found, as shown in Fig. \ref{fig:tri}. On the
square lattice, there are two symmetric NN-RVB spin liquids and
four symmetric NNN-RVB states, as is shown in Fig. \ref{fig:squ}.

For these symmetric RVB spin liquids, it is not known {\it a
priori} whether various correlations decay in power law or
exponentially. Generically, numerical MC simulations
are capable of revealing those
features\cite{liang-88}. The correlations of a physical quantity
$O$ are given by \bea
\avg{O_i O_j}&=&\frac{\bra{\psi_\textrm{RVB}}O_i O_j\ket{\psi_\textrm{RVB}}} {\braket{\psi_\textrm{RVB}}{\psi_\textrm{RVB}}},\nn\\
&=&\frac{\sum_{c,c'}\braket{c}{c'} \left[\frac{\bra{c}O_i
O_j\ket{c'}}{\braket{c}{c'}}\right] }{\sum_{c,c'}\braket{c}{c'}},
\eea where $\braket{c}{c'}$ ($|\braket{c}{c'}|$) can be taken as
statistical weight in MC simulations when they are positive
(negative). For instance, for the square lattice NN RVB state with
$\{\phi_p =0\}$, $\langle c | c' \rangle >0$ for any $c$ and $c'$,
on which large-scale loop algorithm MC
simulations\cite{sandvik-06} were performed recently, reporting
evidence that this RVB state is critical with power-law decaying
dimer correlations\cite{albuquerque-10,tang-11}.

However, it is impossible to choose  $\braket{c}{c'}\geq 0$ for
all $c$ and $c'$ for NN RVB states on non-bipartite lattices (e.g.
the triangular lattice) or NNN RVB states on bipartite lattices
(e.g. the square lattice). Such states are examples of frustrated
RVB wave functions defined as ones whose valence bonds form
non-bipartite graphs.   It is clear that loop-algorithm MC
simulations on frustrated RVB states suffer from the minus sign
problem in the variational level. In the following, we shall show
that the RVB states in \Eq{eq:brvb} can be exactly mapped into
Gutzwiller projected BCS states, which are friendly to MC
simulations without the minus sign problem.

{\bf Projected BCS states}: It was known that the variational
Monte Carlo method has been quite successful in simulating
Gutzwiller projected BCS wave functions. We consider the following
projected BCS wave functions: \bea\label{eq4}
\ket{\psi_\textrm{p-BCS}}=\mathcal{P}_G \exp\left[\sum_{(ij)}
g_{ij} (c^\dag_{i\A}c^\dag_{j\V}-c^\dag_{i\V}c^\dag_{j\A})
\right]\ket{0}, \eea where $c^\dag_{i\sigma}$ are electron
creation operators, $\ket{0}$ is the vacuum, $\mathcal{P}_G$ is
the Gutzwiller projection onto singly occupied states, and
$g_{ij}=g_{ji}$ which are assumed to be real. A similar gauge
symmetry exists for the projected BCS wave functions: the wave
function is invariant, up to a phase, under the transformations:
$g_{jk}\to \exp(i\alpha^f_j) g_{jk} \exp(i\alpha^f_k)$. For time
reversal invariant states with real $g_{jk}$, we define fermionic
fluxes $\phi^f_p$ through $\prod_{(jk)\in p}
\sgn(g_{jk})=\exp(i\phi^f_p)$ with $\phi^f_p=0,\pi$ (mod $2\pi$).
Similarly, the flux pattern $\{\phi^f_p\}$ and
$\{\phi^f_p+(-1)^{n_p}\pi\}$ are equivalent through the gauge
transformation $\exp(i\alpha^f_j) =i$ on all sites $j$. As shown
in details in the Appendix, we obtain
$\ket{\psi_\textrm{p-BCS}}=\ket{\psi_\textrm{RVB}}$ when the
following conditions are satisfied: \bea
|g_{jk}|=|f_{jk}|,~~\phi_p=\phi^f_p+\pi, \eea where $p$ labels all
possible elementary plaquettes.

\begin{table}
\begin{tabular}{|c|c|c|c|c|}
\hline
Kagome NN-RVB state &  A  &  B  &  C  &  D\\
\hline
$\xi_s$ & 0.6 & 0.6 &0.6 & 0.7\\
\hline
$\xi_d$ & 1.2 & 1.0　& 1.0 &0.9\\
\hline $E/J$　& $-0.393$ & $-0.36$ & $-0.357$ & $-0.386$ \\
\hline
\end{tabular}
\caption{The spin ($\xi_s$) and dimer ($\xi_d$) correlation lengths of the four symmetric
states on the Kagome lattice shown in Fig. \ref{fig:kag}. Here $E$
labels the variational energy per site of those symmetric states
for the Kagome NN antiferromagnetic Heisenberg model
$H=J\sum\limits_{\avg{ij}}\mathbf{S}_{i}\cdot\mathbf{S}_{j}$.}
\label{tab:kag}
\end{table}

{\bf Ground state correlations}: We have investigated a number of
short-range frustrated RVB states by performing MC simulations on
their corresponding projected BCS wave functions. In our numerical
calculations, we mainly focus on spin as well as dimer
correlations and study whether they fall off in power law or
exponentially at large distance. The spin and dimer correlations
are defined as follow: \bea\label{corr}
\!S(\vec l)\!\!&=&\!\!\avg{\mathbf{S}_{\vec r_i} \cdot\mathbf{S}_{\vec r_i+\vec l}},\nn\\
\!D_{\alpha}(\vec l)\!\!&=&\!\!\avg{\!\big(\mathbf{S}_{\vec r_i}
\cdot\mathbf{S}_{\vec r_i+\mathbf{e}_\alpha}\big)\big(\mathbf{S}_{\vec r_i+\vec
l} \cdot\mathbf{S}_{\vec r_i+\vec
l+\mathbf{e}_\alpha}\big)\!}\!-\!\avg{\mathbf{S}_{\vec r_i}
\cdot\mathbf{S}_{\vec r_i+\mathbf{e}_\alpha}\!}^2,~~~ \nn\eea
where
$\mathbf{e}_\alpha$ labels lattice vectors.

We have studied all the four symmetric NNN-RVB states
on the square lattice, the four symmetric NN-RVB
states on the Kagome lattice, and the two symmetric
NN-RVB states on the triangular lattice. As discussed in details
later, for all these frustrated RVB states our MC simulations
convincingly show that their spin and dimer correlations fall off
exponentially with correlation length in the order of one lattice
constant, indicating that they are all gapped $Z_2$ quantum spin
liquids. (Note that different symmetric spin liquid states on the
same lattice may be distinguished by numerically computing local
correlations.) We conjecture that our results are generic: all
frustrated short-range RVB states in 2D are fully gapped.

\begin{figure}[tb]
\includegraphics[scale=0.28]{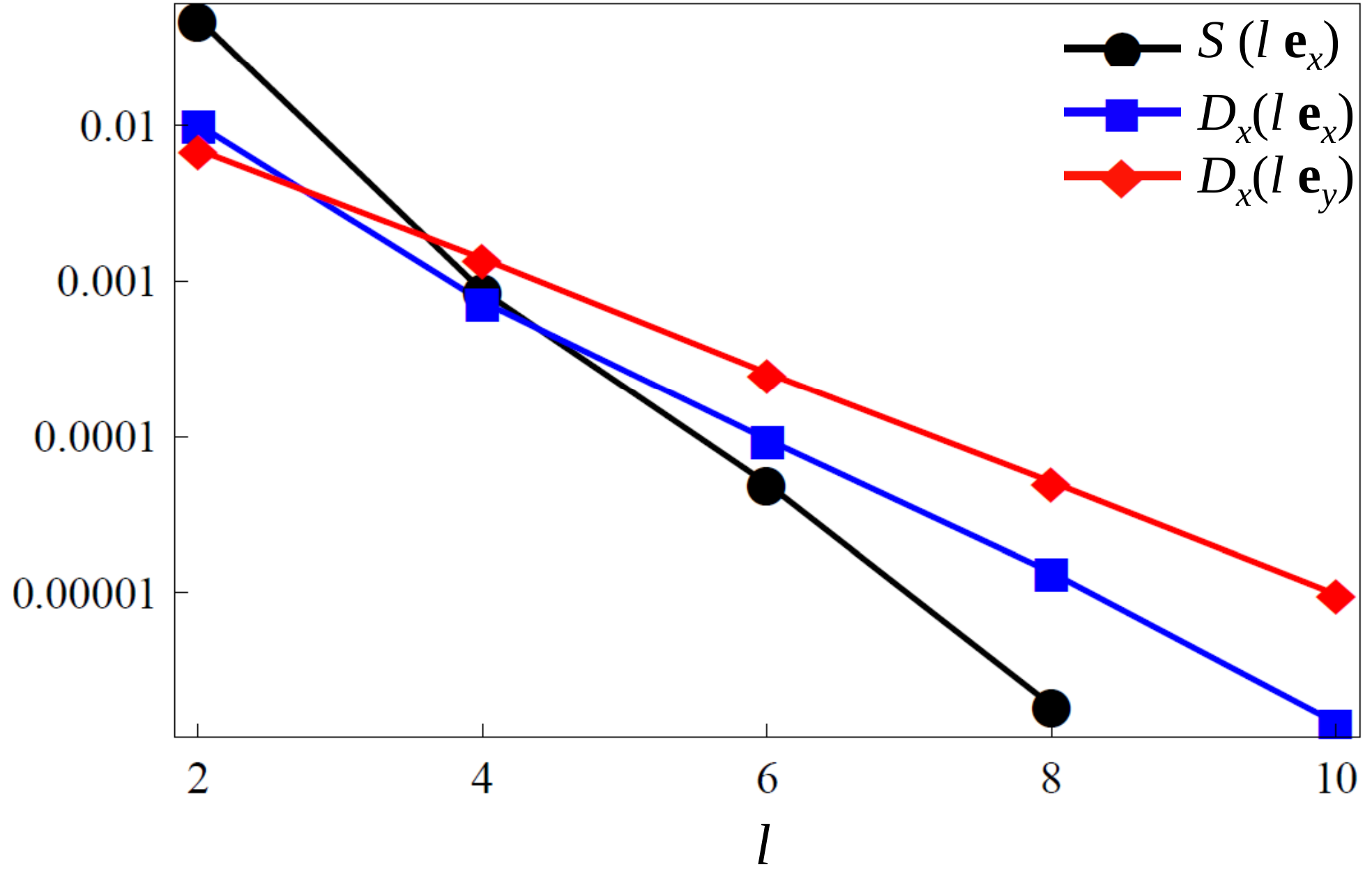}
\caption{The spin and dimer correlations as a function of distance in the NN-RVB state (A) shown in Fig. \ref{fig:kag}.} \label{fig:kag2}
\end{figure}

{\it The Kagome lattice}: According to the flux pattern (either
$\{\phi_p\}$ or $\{\phi^f_p\}$) on the Kagome lattice, it is
straightforward to show that there are four symmetric NN-RVB spin
liquids, as shown in Fig. \ref{fig:kag}. We have computed the spin
and dimer correlations for all the four symmetric NN-RVB spin
liquids. In Fig. \ref{fig:kag2}, we plot the spin and dimer
correlations as a function of distance ($l$) in one of those
NN-RVB symmetric states, {\it i.e.} the
counterclockwise-counterclockwise NN-RVB state (A) shown in Fig.
\ref{fig:kag}. The MC calculations are carried out on a lattice
with $18\times18\times 3$ sites and with periodic boundary
conditions. It is remarkable that the correlations decay extremely
fast. From Fig. \ref{fig:kag2}, it is clear that both spin and
dimer correlations decay exponentially with distance. The spin
correlation length $\xi_s$ for this NN-RVB state is about 0.6
lattice constants. The dimer correlation lengths $\xi_{d,x}$ and
$\xi_{d,y}$ for $D_{x}(l\mathbf{e}_x)$ and $D_{x}(l\mathbf{e}_y)$
are nearly equal, which are about 1.2 lattice constants. The spin
and dimer correlation lengths [$\xi_s$ and $\xi_d\equiv
(\xi_{d,x}+\xi_{d,y})/2$, respectively] for all the four symmetric
NN-RVB spin liquids are listed in Table \ref{tab:kag}. The
correlation lengths are all in the order of one lattice constant,
indicating that they are fully gapped $Z_2$ quantum spin liquids.
This is consistent with the recent density matrix renormalization
group evidence that the ground state of the Kagome Heisenberg
antiferromagnet is a fully gapped quantum spin
liquid\cite{yan-11,depenbrock-12} with correlation lengths of
about one lattice spacing\cite{depenbrock-12}.

To get a sense of which of the above states is the best
variational wave function for the Kagome antiferromagnet described
by $H=J\sum_{\avg{ij}}\mathbf{S}_{i}\cdot\mathbf{S}_{j}$, we
compute their variational energy per site as is shown in Table
\ref{tab:kag}. Among the four symmetric NN-RVB spin-liquid states,
the counterclockwise-counterclockwise NN-RVB state has the lowest
energy for the Kagome antiferromagnet, which is $-0.393J$ per
site. This energy still differs from the density matrix
renormalization group result, indicating that longer-range valence
bonds are important in describing the Kagome antiferromagnet.

{\it The triangular lattice}: It turns out that there are only two
symmetric NN-RVB spin liquid states on the triangular lattice,
whose flux patterns $\{\phi_p\}$ are shown in Fig. \ref{fig:tri}.
For both states, the spin and dimer correlations decay
exponentially with distance, with $\xi_s=0.7$ and $\xi_d=1.0$ for
the state (A) shown in Fig. \ref{fig:tri} and $\xi_s=1.0$ and
$\xi_d=1.6$ for the state (B) shown in Fig. \ref{fig:tri}. Both
NN-RVB states are then gapped $Z_2$ spin liquids. The correlation
lengths on the triangular lattice are somewhat longer than those
on the Kagome lattice, which is expected due to more geometric
frustrations in the Kagome lattice.

\begin{table}[t]
\begin{tabular}{|c|c|c|c|c|}
 \hline square NNN-RVB state & A & B & C & D\\ \hline
$\xi_s$ & 1.1 & 0.8 &0.7 & 0.6\\ \hline
$\xi_{d,\textrm{NN}}$　　& 1.2 &
1.6　& 1.3 &1.4\\ \hline
$\xi_{d,\textrm{NNN}}$　　& 0.6 & 0.6　& 0.6 &0.5\\
\hline $E/J_1$　& $-0.344$ & $-0.219$　& $-0.237$ &$-0.239$ \\
\hline
\end{tabular}
\caption{The correlation lengths and variational energies of the four symmetric NNN-RVB states on the square lattice.}
\label{tab:squ}
\end{table}

{\it The square lattice}: It was shown recently that the
unfrustrated NN-RVB state on the square lattice is a critical
state with power-law dimer correlations even though its spin
excitations are gapped\cite{albuquerque-10,tang-11,fendley-10}. To
have a fully gapped spin-liquid phase on this lattice, it is
necessary to include frustration in short-range RVB states. In
this Letter, we consider to include NNN valence bonds, which is
partly motivated by a recent study establishing that fully gapped
spin-liquid ground states are realized in the
generalized\cite{yao-11} quantum dimer models\cite{kivelson-88}
with NN and NNN dimers on the square lattice. (Note that the
nature of NNN RVB states in \Eq{eq:brvb} without the factor
$\left(-1\right)^{\delta_{c}}$ remains unknown due to the lack of
mapping between them and projected BCS states.)

From the flux pattern $\{\phi_p\}$, we have identified four
symmetric NNN-RVB states on the square lattice, as shown in Fig.
\ref{fig:squ}(b). For each of these four states, there is an
additional parameter labeling the wave function, namely the ratio
$\gamma\equiv |f_\textrm{NNN}/f_\textrm{NN}|$. We take $\gamma=1$
in our MC simulations. For $\gamma=1$ both spin and dimer
correlations decay exponentially with distance. The correlation
lengths are listed in Table \ref{tab:squ}, where
$\xi_{d,\textrm{NN}}$ and $\xi_{d,\textrm{NNN}}$ mean the
correlation lengths of NN and NNN dimers, respectively. At
$\gamma=1$ (more generally $\gamma>0$), we conclude that the four
symmetric NNN-RVB states are fully gapped $Z_2$ spin-liquids,
which is consistent with the recent numerical evidence of fully
gapped spin-liquids in the $J_1$-$J_2$ square Heisenberg
antiferromagnet\cite{jiang-11,wang-11}. The variational energies
of these NNN-RVB states in unit of $J_1$ for the $J_1$-$J_2$
square Heisenberg model with $J_2=J_1/2$ are shown in Table
\ref{tab:squ}.

\begin{figure}
\subfigure{\includegraphics[scale=0.15]{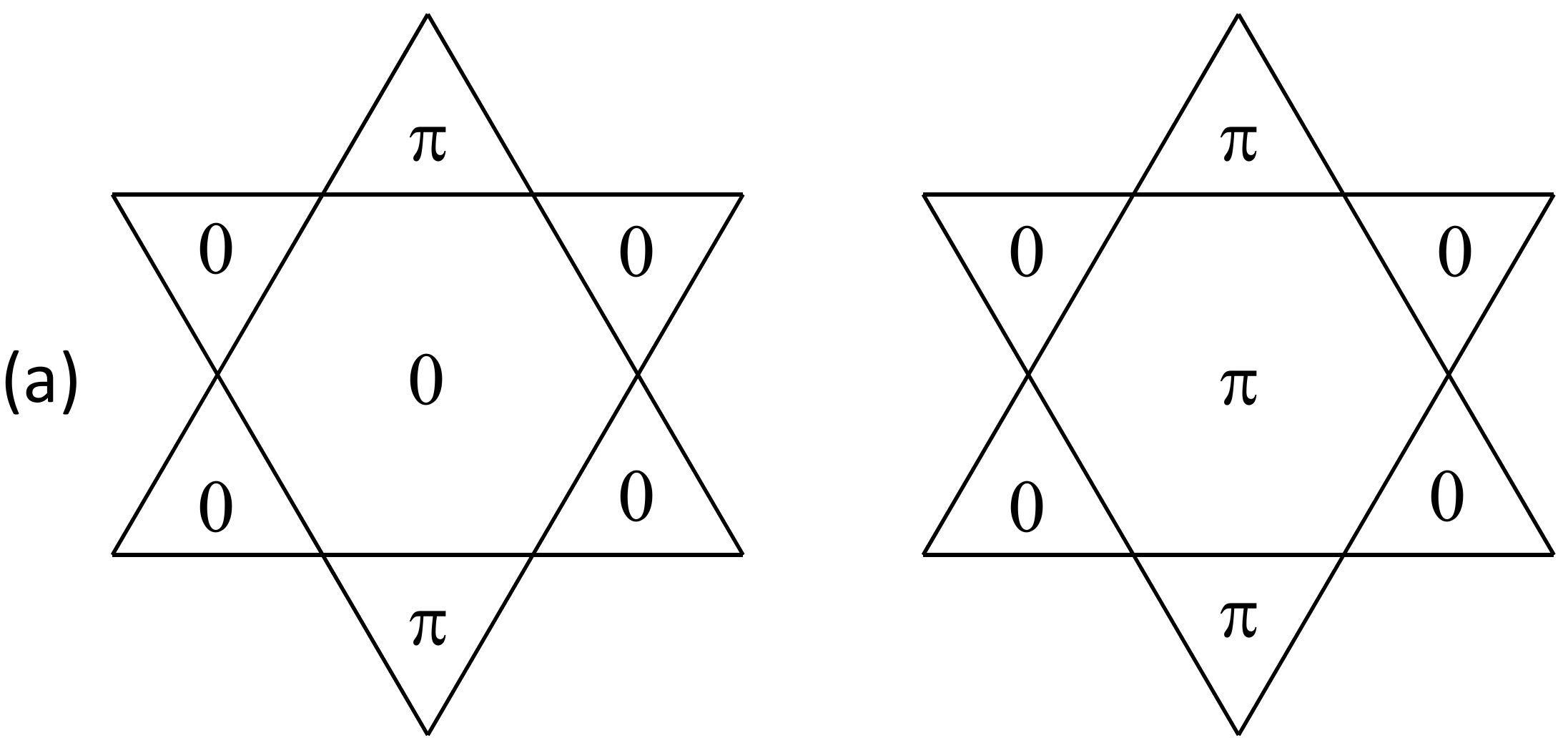}}~~
\subfigure{\includegraphics[scale=0.15]{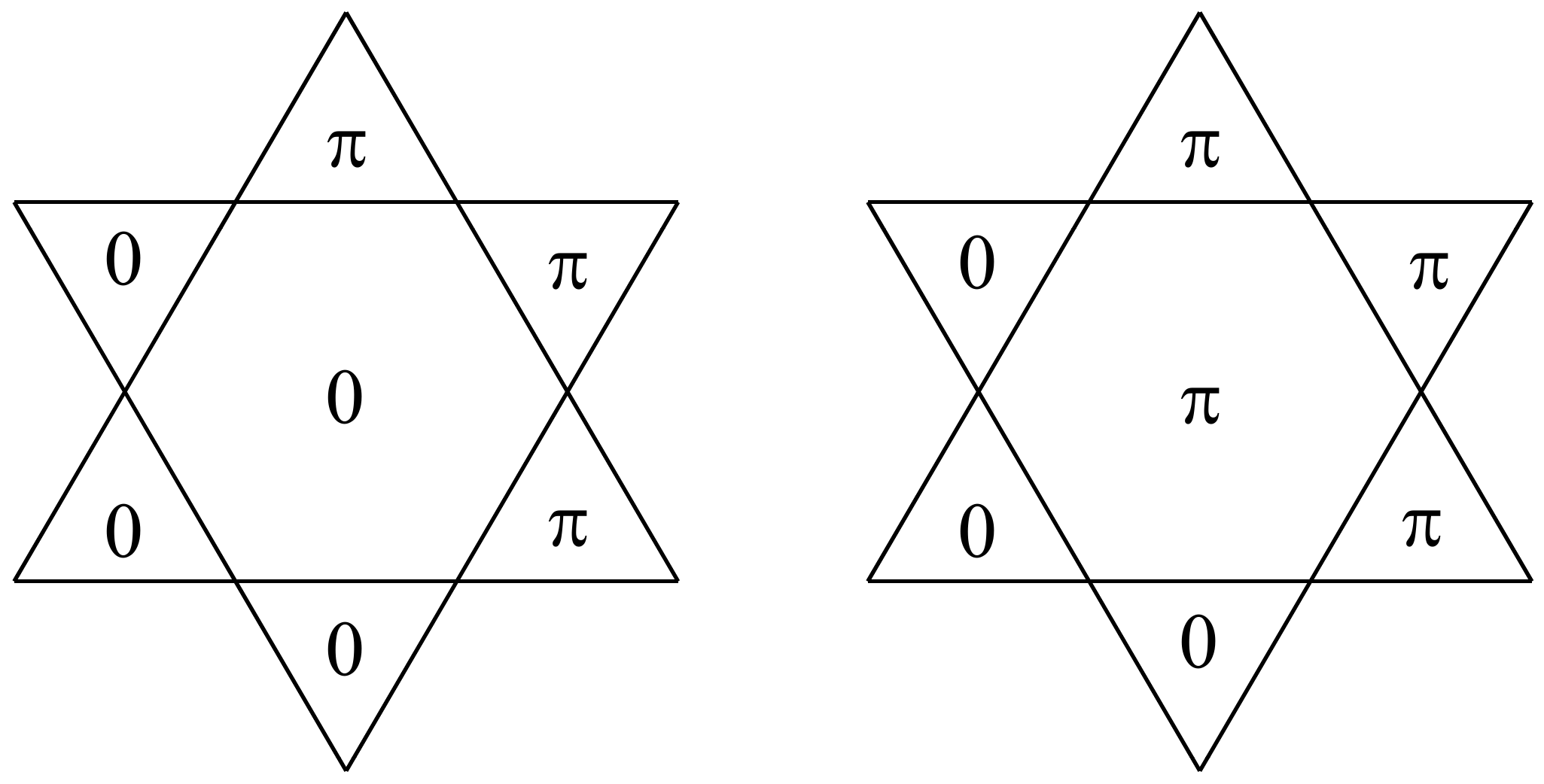}}\\
\subfigure{\includegraphics[scale=0.20]{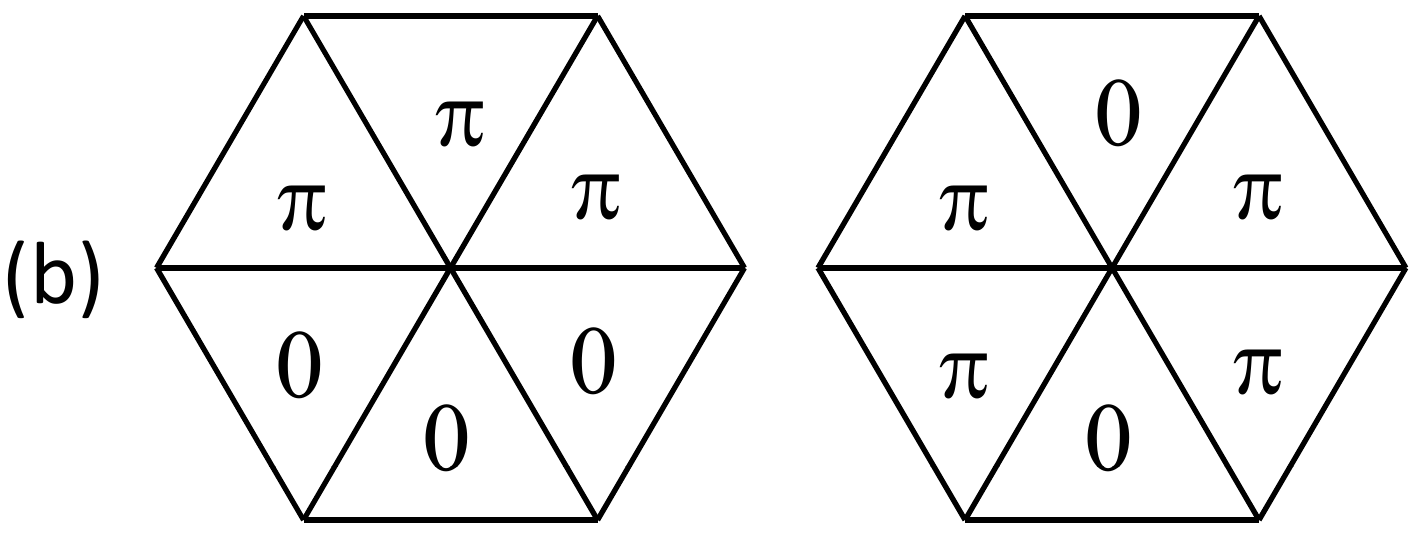}}\\
\subfigure{\includegraphics[scale=0.22]{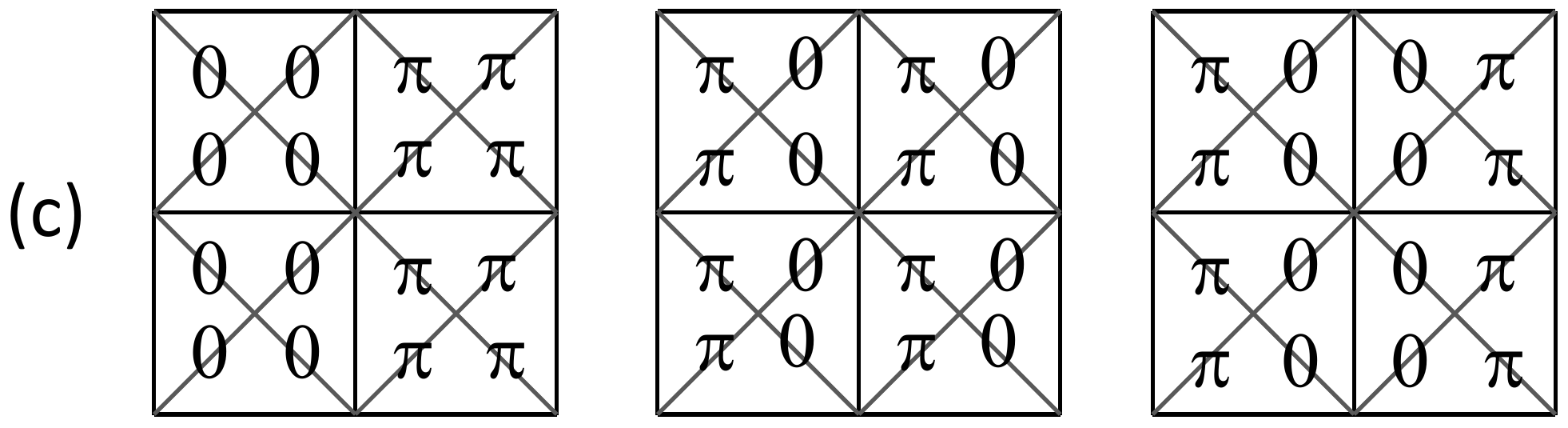}}~
\subfigure{\includegraphics[scale=0.22]{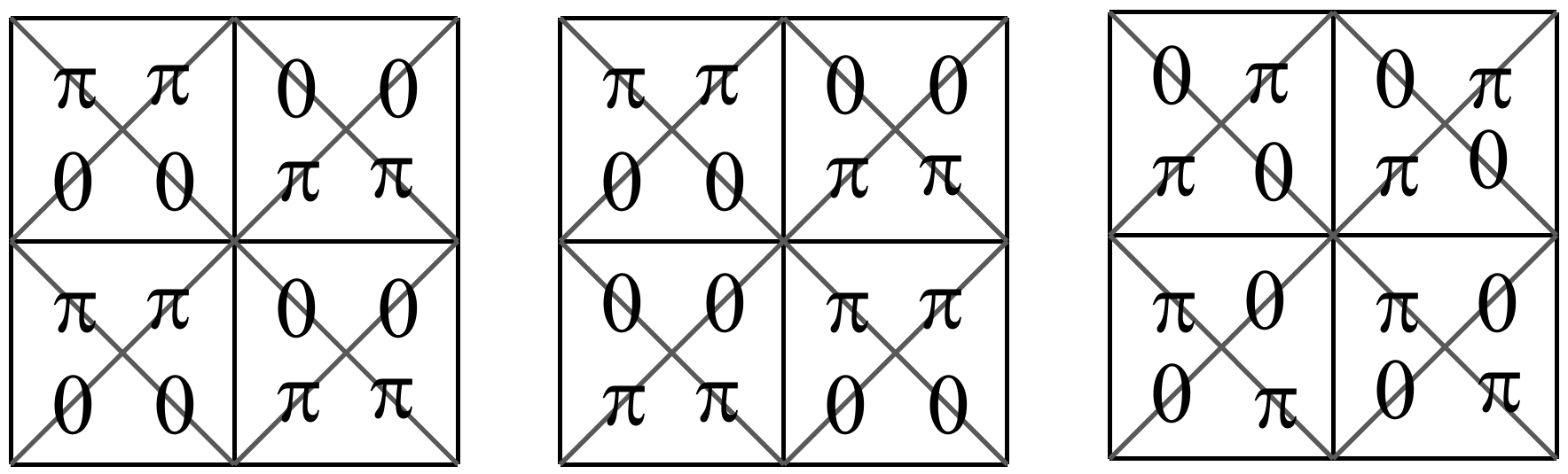}}
\caption{The flux patterns $\{\phi_p\}$ of (a) the four nematic NN-RVB states on the Kagome lattice, (b) the two nematic NN-RVB states on the triangular lattice, and (c) the six nematic NNN-RVB states on square lattice.
}
\label{fig:nem}
\end{figure}

{\bf Nematic RVB spin-liquids}: We have studied fully symmetric
short-range RVB spin-liquids on various lattices. An interesting
question is whether short-range RVB states could be nematic
spin-liquids which are translationally invariant but break lattice
point group symmetry. The answer is yes. On the Kagome lattice, we
identified that there are only four NN-RVB states which are
nematic spin-liquids, as shown in Fig. \ref{fig:nem}(a). On the
triangular lattice, there are two nematic NN-RVB states, as shown
in Fig. \ref{fig:nem}(b). Our MC simulations show that the
correlation functions of spins and dimers in these states decay
exponentially with distance but the $C_{6v}$ symmetry of both
lattices is broken. They are fully gapped nematic spin-liquids, in
contrast with gapless nematic spin-liquids on the triangular
lattice studied in Ref. \cite{senthil}.

On the square lattice, we found six nematic NNN-RVB spin-liquid
states, which are shown in Fig. \ref{fig:nem}(c). These
spin-liquid states keep the translational symmetry but breaks the
$C_{4v}$ rotational symmetry of the square lattice. Again, our MC
simulations show that they are fully gapped nematic spin-liquids.

{\bf Concluding discussions}:
On the square (or honeycomb) lattice,
there are two kinds of symmetric NN-RVB spin-liquids. The $\{\phi_p=0\}$ NN-RVB state
is unfrustrated with power-law decaying dimer correlations\cite{albuquerque-10,tang-11}.
 For the $\{\phi_p=\pi\}$ NN-RVB state, our MC simulations on the square
 lattice with $40\times 40$ sites implies that its dimer correlation
 decays in power law even though we need further studies on systems with larger sizes
 to determine the power exponent accurately.
 [Here the power-law decaying dimer correlations are expected since bipartite RVB states are effectively described by an emergent U(1) gauge theory.].

A recent loop-algorithm MC study shows that the unfrustrated
NN-RVB states on the cubic and diamond lattices show magnetic
long-range order\cite{moessner-12}. Properties of NN-RVB states on
3D frustrated lattices remain unknown partly due to the minus sign
problem in loop-algorithm MC simulations. We can generalize the
mapping between short-range bosonic RVB states and projected BCS
states discussed in the present Letter to three dimensions.
Consequently, we can solve the minus sign problem for a class of
frustrated short-range RVB states, which is a significant step
towards understanding the nature of short-range frustrated RVB
states in 3D.

{\bf Acknowledgments}: We are grateful to Zheng-Yu Weng and Tao
Xiang for sharing computing resources and thank Steve Kivelson and
Tao Li for helpful discussions. This work is supported in part by
the NSFC under Grants No. 10704008 and No. 11274041 (F.Y.), and by
Tsinghua Startup Funds and NSF Grant No. DMR-0904264 (H.Y.).

{\it Note added}: After the completion of the present Letter, we notice Ref. \cite{seidel-12} where similar results were obtained for a subset of NN RVB states classified in the present Letter.

\begin{widetext}

\subsection{Supplemental Material: proof of exact mapping between bosonic RVB and projected BCS states}
Now, we shall prove the mapping between a class of short-range RVB states and Gutzwiller projected BCS wave functions\cite{sorella-06}. The former is given by
\bea
\ket{\psi_\textrm{RVB}}&=&\sum_c (-1)^{\delta_c} \prod_{(ij)\in c} f_{ij}\ket{ij},\\
&=&\sum_c (-1)^{\delta_c} \left[\prod_{(ij)\in c} f_{ij}(S^-_j-S^-_i)\right] \ket{\A\A\cdots\A},\\
&=&\sum_c (-1)^{\delta_c}(-1)^{p_c} \left[\prod_{(ij)\in c} f_{ij}(S^-_j-S^-_i)c^\dag_{i\A} c^\dag_{j\A}\right]\ket{0},\\
&=&\sum_c (-1)^{\delta_c}(-1)^{p_c} \left[\prod_{(ij)\in c}f_{ij} (c^\dag_{i\A} c^\dag_{j\V}-c^\dag_{i\V}c^\dag_{j\A})\right]\ket{0},
\eea
where $p_c$ is the signature of $c$ (some ordering in $(ij)$ for each $c$ is implicitly assumed.)

Let's first consider NN-RVB states, for which $\delta_c=0$ for any
$c$. The NN-RVB states are given by \bea\label{eq1}
\ket{\psi_\textrm{NN-RVB}}=\sum_c  \left[(-1)^{p_c}\prod_{(ij)\in
c} f_{ij} (c^\dag_{i\A}
c^\dag_{j\V}-c^\dag_{i\V}c^\dag_{j\A})\right] \ket{0}\equiv
\sum_c\ket{c}. \eea The corresponding projected BCS
wave function \Eq{eq4} with $|g_{ij}|=|f_{ij}|$ can be expanded as follows:
\bea
\ket{\psi_\textrm{p-BCS}}&=&\mathcal{P}_G\frac{1}{\left(\frac{N}{2}\right)!}
\left[\sum_{(ij)} g_{ij}
(c^\dag_{i\A}c^\dag_{j\V}-c^\dag_{i\V}c^\dag_{j\A})
\right]^{\frac{N}{2}}\ket{0},\nonumber\\ &=&\sum_{c}
\left[\prod_{(ij)\in c} g_{ij}(c^\dag_{i\A}
c^\dag_{j\V}-c^\dag_{i\V}c^\dag_{j\A})\right] \ket{0}\equiv\sum_c
\ket{c}_F. \eea
Notice that the second step above is a consequence
of the no-double occupance projection, which excludes the
possibility that one site takes part in two valence-bonds.

Suppose $\ket{c_0}=\ket{c_0}_F$ for some $c_0$ by choosing
appropriate signs of $g_{ij}$ for $(ij)\in c_0$, which can always be done. We have \bea (-1)^{p_{c_0}}\prod_{(ij)\in
c_0} \sgn(f_{ij}) = \prod_{(ij)\in c_0}\sgn(g_{ij}). \eea It is
clear that two wave functions $\ket{\psi_\textrm{NN-RVB}}$ and
$\ket{\psi_\textrm{p-BCS}}$ are equal if \bea\label{eq15}
(-1)^{p_c}\prod_{(ij)\in c} \sgn(f_{ij}) = \prod_{(ij)\in
c}\sgn(g_{ij}) \eea for any $c$. \Eq{eq15} is equivalent to
\bea\label{eq16} (-1)^{p_c+p_{c_0}}\prod_{(ij)\in c+c_0}
\sgn(f_{ij}) = \prod_{(ij)\in c+c_0}\sgn(g_{ij}), \eea where
$c+c_0$ represents a transition graph obtained by superposing two
valence bonds configurations $c$ and $c'$ on the same lattice. The
transition graph consists of even-length loops of valence bonds.
It is clear that \Eq{eq16} is satisfied if \bea\label{eq17}
\prod_{(ij)\in p} \sgn(g_{ij}) =-\prod^{cc}_{(ij)\in p}
\sgn(f_{ij}) \eea for every even-length plaquette $p$. The minus
sign in \Eq{eq17} comes from the signature $(-1)^{p_c+p_{c_0}}$
when the ordering of $(ij)$ in $c+c_0$ is taken in a
counterclockwise fashion. For \Eq{eq17} to be true, it is
sufficient to have \bea\label{eq18} \phi_p=\phi^f_p+\pi \eea for
all possible elementary plaquettes $p$, which could be even or
odd-length. Note that open boundary conditions of the lattice in
question is implicitly assumed here. For lattices with periodic
boundary conditions in both directions, four projected BCS wave
functions are needed to form a linear superposition in order to
represent a short-range RVB state\cite{yao-11,sorella-09}.
Nonetheless, the boundary conditions of lattices do not
qualitatively affect the large-distance correlation functions of
local physical quantities.

Now, we consider the case where both NN and NNN valence bonds are allowed in valence bond configurations and the RVB states are called NNN-RVB states. It is clear that each cross generates a twist from the configuration without that cross, which results in the factor $(-1)$ from  the signature $(-1)^{p_c}$. The factor $(-1)$ from the signature $(-1)^{p_c}$ can be compensated by the factor $(-1)^{n_c}$. This proves that the NNN-RVB states can also be mapped exactly into projected BCS wave functions with pairing between NN and NNN sites.

\end{widetext}

\begin{thebibliography}{33}
\bibitem{note1} The classification here might be closely related to studies based on the projective symmetry group in Ref. \cite{wen-psg,fa}.
\bibitem{note2} Note that the definition of fluxes above for
time reversal invariant RVB states with real $f_{ij}$ is
consistent with the more general one on the even-length plaquettes
studied in Refs. \cite{fa,oleg}.
\end{thebibliography}
\end{document}